\documentclass[onecolumn,preprint3]{aastex63}
\usepackage{lineno}
\usepackage{bbding}
\usepackage{amssymb,amsmath,bm}
\usepackage{rotating}
\usepackage{hyperref}
\usepackage{bm}
\hypersetup{
  colorlinks      = {true},
  linkcolor       = {blue},
  citecolor       = {blue},
  urlcolor        = {blue},
}

\shorttitle{}
\shortauthors{}
\begin{document}

\title{Joint Constraints on the Hubble Constant, Spatial Curvature, and Sound Horizon from the Late-time Universe with Cosmography}

\author{Kaituo Zhang}
\affiliation{Department of Physics, Anhui Normal University, Wuhu, Anhui 241000, China}

\author{Tianyao Zhou}
\affiliation{Department of Physics, Anhui Normal University, Wuhu, Anhui 241000, China}

\author{Bing Xu}
\affiliation{School of Electrical and Electronic Engineering, Anhui Science and Technology University, Bengbu, Anhui 233030, China}
\email{xub@ahstu.edu.cn}

\author{Qihong Huang}
\affiliation{School of Physics and Electronic Science, Zunyi Normal University, Zunyi 563006, Guizhou, China}

\author{Yangsheng Yuan}
\affiliation{Shandong Provincial Engineering and Technical Center of Light Manipulations and Shandong Provincial Key Laboratory of Optics and Photonic Device, School of Physics and Electronics, Shandong Normal University, Jinan 250014, China;}
\affiliation{Collaborative Innovation Center of Light Manipulation and Applications, Shandong Normal University, Jinan 250358, China; }
\affiliation{Joint Research Center of Light Manipulation Science and Photonic Integrated Chip of East China Normal University and Shandong Normal University, East China Normal University, Shanghai 200241, China;}

\begin{abstract}
In this paper, using the latest Pantheon+ sample of Type Ia supernovae (SNe Ia), Baryon Acoustic Oscillation (BAO) measurements, and observational Hubble data (OHD), we carry out a joint constraint on the Hubble constant $H_0$, the spatial curvature $\Omega_{\rm K}$, and the sound horizon at the end of drag epoch $r_{\rm d}$. To be model-independent, four cosmography models, i.e., the Taylor series in terms of redshift $y_1=z/(1+z)$, $y_2=\arctan(z)$, $y_3=\ln(1+z)$, and the Pad\'e approximants, are used without the assumption of flat Universe. The results show that the $H_0$ is anti-correlated with $\Omega_{\rm K}$ and $r_{\rm d}$, indicating smaller $\Omega_{\rm K}$ or $r_{\rm d}$ would be helpful in alleviating the Hubble tension. And the values of $H_0$ and $r_{\rm d}$ are consistent with the estimate derived from the Planck Cosmic Microwave Background (CMB) data based on the flat $\Lambda$CDM model, but $H_0$ is in 2.3$\sim$3.0$\sigma$ tension with that obtained by \cite{Riess2022} in all these cosmographic approaches. Meanwhile, a flat Universe is preferred by the present observations under all approximations except the third order of $y_1$ and $y_2$ of the Taylor series. Furthermore, according to the values of the Bayesian evidence, we found that the flat $\Lambda$CDM remains to be the most favored model by the joint datasets, and the Pad\'e approximant of order (2,2), the third order of $y_3$ and $y_1$ are the top three cosmographic expansions that fit the datasets best, while the Taylor series in terms of $y_2$ are essentially ruled out.
\end{abstract}

\section{Introduction}\label{S1}

In modern cosmology, the Hubble tension has become one of the most prominent issues. The Hubble constant $H_0$, as a fundamental cosmological parameter, represents the expansion rate of the present Universe, which can be obtained by both global and local measurements, i.e., based on the standard cosmological model($\Lambda$CDM), Planck Collaboration inferred $H_0 = 67.4\pm0.5$ km $\mathrm{s}^{-1} \mathrm{Mpc}^{-1}$ (hereafter P18)~\citep{Aghanim2020} from Planck satellite measurements of Cosmic Microwave Background (CMB) temperature and polarization anisotropies, while, Supernovae and $H_0$ for the Equation of State of dark energy (SH0ES) collaboration found $H_0 = 73.04\pm1.04$ km $\mathrm{s}^{-1} \mathrm{Mpc}^{-1}$ (hereafter R22)~\citep{Riess2022} via local measurements from type Ia supernovae (SNe Ia) calibrated by the distance ladder without any cosmological model. The 5$\sigma$ tension between these two independent $H_0$ estimates could be pointing towards new physics beyond the standard model or residual systematics~\citep{Freedman2017,Feeney2018,Di Valentino2018,Handley2021}. To alleviate the discrepancy, fundamental physics beyond $\Lambda$CDM are investigated, such as time-dependent dark energy equation of state~\citep{Huang&Wang2016,Di Valentino2016,Di Valentino2021,Zhao2017,Miao&Huang2018,Yang2019,Poulin2019,Vagnozzi2021,Colgain2021,Yang2023}, modified gravity~\citep{Capozziello2003,Nunes2018,Farrugia2021,Koussour2022,Sultana2022}, and additional relativistic particles, see~\citet{Kumar&Nunes2016,Xu&Huang2018,Carneiro2019,Pandey2020,D'Eramo2022}. Alternately, new approaches to determine the Hubble constant from local direct measurements are proposed. For example, through time-delay cosmography, the $H_0$ Lenses in COSMOGRAIL’s Wellspring (H0LiCOW) Collaboration~\citep{Suyu2017} measured the Hubble constant from strong gravitational lens systems with time delays between the multiple images. Using detected gravitational waves (GW) as a ‘standard siren’,  binary neutron-star (BNS) system detection GW170817 and subsequent observations in the electromagnetic (EM) domain provide another independent method of measuring the Hubble constant~\citep{Abbott2017}. Adopting revised measurement, \citet{Freedman2020} found $H_0=69.9\pm0.8$($\pm1.1\%$ stat)$\pm1.7$($\pm2.4\%$ sys)km $\mathrm{s}^{-1} \mathrm{Mpc}^{-1}$, which provided one of the most accurate means of measuring the distances to nearby galaxies by the red giant branch method.

However, if the systematic uncertainties are not the main drivers of $H_0$ tension, then the discrepancy coming from the concrete cosmological model assumption can not be ignored. In order to extricate from dependency on a cosmological model and study the expansion of the Universe directly from the observations, various model-independent techniques are used, such as cosmography, the B\'ezier parametric curve, the Parameterization based on cosmic Age (PAge), the Gaussian process (GP) method and so on~\citep{Wojtak&Agnello2019,Capozziello2019,Yang2020,Zhang&Huang2021,Cai2022a,Cai2022b,Hu&Wang2022,Jalilvand&Mehrabi2022,Liu2022}. For examples, \citet{Zhang&Huang2021} reconstructed $H(z)$ in cubic expansion and polynomial expansion respectively, and constrained $H_0$ and $r_{\rm d}$ with the joint data of SNe Ia, Baryon Acoustic Oscillation (BAO) measurements, Observational Hubble Data (OHD), and GW data, and found the $H_0$ value is in 2.4-2.6$\sigma$ tension with SH0ES 2019~\citep{Riess2019}. \citet{Cai2022a,Cai2022b} applied the PAge approximation to consistently use the OHD and the late-time matter perturbation growth data at high redshifts, and found the Hubble tension can't be solved by introducing the new physics at the late time beyond the $\Lambda$CDM model. \citet{Hu&Wang2022} investigated the redshift-evolution of $H_0$ with 36 Hubble parameter $H(z)$ data based on the GP method, and found there was a late-time transition of $H_0$ which effectively alleviated the Hubble crisis by $70\%$. It should be pointed out that these works are all taken under the assumption of a spatially flat Universe. However, there is still one observational probe that is in agreement with a negative curvature, sparking the debate about the flatness of the Universe. From the combination of CMB temperature and polarization power spectra, the constraint on curvature suggests a closed Universe at more than three standard deviations ($\Omega_{\rm K}=-0.044^{+0.018}_{-0.015}$)~\citep{Aghanim2020}. Whereas once the Planck CMB data are combined with the BAO measurements, a flat Universe is preferred with a high precision $\Omega_{\rm K}= 0.001\pm0.002$~\citep{Aghanim2020}, indicating there is a discrepancy between CMB and BAO data for the constraint on $\Omega_{\rm K}$, namely ``curvature tension''~\citep{Handley2021}. As allowing for spatial curvature may significantly affect constraints on cosmological parameters~\citep{Dossett&Ishak2021}, the curvature parameter also needs to be considered when exploring Hubble tension. For instance, it is found that a negative curvature preferred by P18 will exacerbate $H_0$ tension~\citep{Di Valentino2020,Di Valentino2021}, while~\citet{Wang2021} and~\citet{Cao2022} found that a spatially flat Universe is favored using the mock GW data in combination with the Hubble parameter or strong gravitational lensing time delay (SGLTD) data respectively. Therefore, due to the significant influence of spatial curvature on the $H_0$ measurement and the inconsistency of the spatial curvature measurement, it is very necessary for us to measure $H_0$ with various observations under the assumption of a non-flat Universe.

Additionally,  since that the value of $H_0$ obtained from the Planck CMB data is directly tied to the sound horizon at last scattering, which is closely related to the sound horizon $r_{\rm d}$ at the baryon decoupling, a number of amendments to the $\Lambda$CDM model have been proposed, aiming to release the Hubble tension by reducing $r_{\rm d}$ and increasing $H_0$~\citep{Karwal&Kamionkowski2016,Buen-Abad2018,Anchordoqui&Bergliaffa2019,Niedermann&Sloth2020,Berghaus&Karwal2020,Archidiacono2020}. However, the most recent works demonstrated that any model which only reduces $r_{\rm d}$ can never
fully resolve the Hubble tension~\citep{Pogosian2020,Jedamzik2021}. In addition, in order to break the measured degeneracy between $r_{\rm d}$ and $H_0$, many works have therefore attempted to measure the $H_0$ and $r_{\rm d}$ by combining the BAO measurements with other late-time observations of Universe. For instance, \citet{Wojtak&Agnello2019} combined BAO with the SGLTD, Joint Light-Curve Analysis (JLA), and the reconstructed $H(z)$ data via the polynomial expansion, and obtained $H_0=72.3\pm6.9$ km $\mathrm{s}^{-1} \mathrm{Mpc}^{-1}$, $r_{\rm d}=139.2\pm13.3$ Mpc. Then ~\citet{Pogosian2020} found $r_{\rm d}=143.7\pm2.7$ Mpc and $H_0=69.6\pm1.8$ km $\mathrm{s}^{-1} \mathrm{Mpc}^{-1}$ by using the latest BAO data along with a $\Omega_{\rm m} h^2$  prior based on the Planck best-ﬁt $\Lambda$CDM model, and similar values were obtained when they combined BAO with the Pantheon supernovae, the Dark Energy Survey Year 1 galaxy weak lensing~\citep{Abbott2018}, Planck/SPTPo1 CMB lensing, and OHD. In addition, \citet{Cai2022b} combined the SNe Ia, BAO and OHD, and obtained $H_0=68.958^{+1.779}_{-1.826}$ km $\mathrm{s}^{-1} \mathrm{Mpc}^{-1}$ and $r_{\rm d}=146.466^{+3.448}_{-3.302}$ Mpc by using the PAge approximation. Although the Hubble tension can be alleviated in these works, it cannot be completely resolved. To reflect the reality of these works taken under the assumption of a spatially flat Universe, it is well worth constraining $H_0$ and $r_{\rm d}$ with leaving $\Omega_{\rm K}$ free by using the latest low-redshift observational data.

In this work, we therefore plan to carry out a joint constraint on $H_0$, $\Omega_{\rm K}$ and $r_{\rm d}$ with the latest late-time observations of Universe including SNe Ia, BAO, and OHD. Specially, the newest Pantheon+ sample of SNe Ia~\citep{Brout2022, Scolnic2022} ranging in redshift from $z$ = 0.001 to 2.26, is used. This sample is made of 1701 light curves of 1550 spectroscopically confirmed SNe Ia and significantly enlarges the origin Pantheon sample size from the addition of multiple cross-calibrated photometric systems of SNe Ia. For the BAO measurements, fourteen latest measurements of $D_{\rm V}(z)/r_{\rm d}$, $D_{\rm M}(z)/r_{\rm d}$ and $D_{\rm H}(z)/r_{\rm d}$ summarized in Ref.~\citep{Alam2021}, covering the redshift range $0.15\leq z \leq2.33$, are used. These data are obtained from final observations of clustering using galaxies, quasars, and Ly$\alpha$ forests from the completed SDSS lineage of experiments in a large-scale structure, composing of data from SDSS, SDSS-II, BOSS and eBOSS. Furthermore, to be model-independent, the well-consolidated approach named cosmography~\citep{Chiba&Nakamura1998,Caldwell&Kamionkowski2004,Visser2004,Visser2005,Visser2015,Capozziello2013,Capozziello2019,Dunsby&Luongo2016,Zhang2017,Yin2019,Li2020} is used, which has attracted lots of attention in the study of the expansion of Universe. The idea of cosmography is to expand the cosmological distances or the Hubble parameter into a Taylor series of redshift $z$, which performs well at low redshifts but encounters the convergence problems in the high-redshift domain~\citep{Cattoen&Visser2007}. To overcome the convergence issues, several improved approaches have been proposed, one of which relies on the use of auxiliary variables~\citep{Cattoen&Visser2007,Aviles2012,Capozziello2020}, and the other expands observables in terms of rational approximations~\citep{Gruber&Luongo2014,Wei2014,Shafieloo2012,Capozziello2018}. Recently, \citet{Li2020} adopted two cosmographic methods, the Taylor series in terms of $y=z/(1+z)$ and Pad\'e polynomials, to investigate the spatial curvature parameter with the dataset including the Pantheon sample of SNe Ia, BAO, and OHD data, and found the $H_0$ tension problem can be slightly relaxed by introducing the spatial curvature parameter. In order to get more robust results, we explore the Hubble tension through the extended cosmographic techniques with Taylor series in terms of $y_1=z/(1+z)$, $y_2=\arctan(z)$, $y_3=\ln(1+z)$ and Pad\'e approximations using the latest data, while leaving $\Omega_{\rm K}$ and $r_{\rm d}$ free at the same time. Meanwhile, a Bayesian approach based on Bayesian evidence is applied to test which cosmographic method is most favored by the observational data.

This paper is organized as follows: In Section~\ref{S2}, we introduce the cosmographic approaches, dataset, and methodology used in this work. In Section~\ref{S3}, our constraint results and analysis are presented. Finally, the conclusions are drawn in Section~\ref{S4}.


\section{Methodology and Data}\label{S2}

The cosmographic approach is an artful combination of kinematic parameters via the Taylor series with the assumption of large-scale homogeneity and isotropy, which can be retained in the Friedmann–Robertson–Walker (FRW) metric,

\begin{equation}
ds^2=-c^2dt^2+a^2(t)\left[\frac{dr^2}{1-Kr^2}+r^2(d\theta^2+\sin^2\theta d\phi^2)\right],
\end{equation}
where $c$ is the speed of light, $K$ is the constant curvature of the three-space of the FRW metric, and $a(t)$ is the scale factor with cosmic time $t$. Considering a photon traveling towards us along a radial path ($ds=0$), we can get

\begin{equation}
c dt=a(t)\frac{dr}{\sqrt{1-Kr^2}}.
\end{equation}
Integrating the equation, the comoving distance can be obtained by

\begin{equation}
d_{\rm c}\equiv a_0\int_{0}^{r_e}\frac{dr}{\sqrt{1-Kr^2}}=-a_0\int_{t_e}^{t_0}\frac{cdt}{a(t)},
\end{equation}
where $t_0$ and $a_0$ are the time and scale factor when the photon was observed from us, and $t_e$ is the time when it was emitted from the source at $r=r_e$. Substituting $1+z=a_0/a$, then the comoving distance becomes

\begin{equation}
d_{\rm c}=\frac{c}{H_0}\int^z_0\frac{dz'}{E(z)},
\label{COMD}
\end{equation}
where $E(z)\equiv{H(z)}/{H_0}$, and the transverse comoving distance~\citep{Hogg1999} can be expressed by

\begin{equation}
D_{\rm M}(z)=\left\{
\begin{aligned}
&\frac{c}{H_0\sqrt{\Omega_{\rm K}}}\sinh\left(\frac{H_0\sqrt{\Omega_{\rm K}}}{c}d_{\rm c}\right), &\quad K<0\\
&d_{\rm c}, &\quad K=0\\
&\frac{c}{H_0\sqrt{-\Omega_{\rm K}}}\sin\left(\frac{H_0\sqrt{-\Omega_{\rm K}}}{c}d_{\rm c}\right), &\quad K>0\\
\end{aligned}
\right.
\label{DM}
\end{equation}
where $\Omega_{\rm K}=-Kc^2/(a_0H_0)^2$ is the present value of spatial curvature parameter. Now define the cosmographic parameters, i.e., Hubble parameter $H$, deceleration $q$, jerk $j$, snap $s$ and lerk $l$ parameters, as follows,
\begin{equation}
\begin{aligned}
&H(t)\equiv \frac{1}{a}\frac{da}{dt},\,\,q(t)\equiv -\frac{1}{aH^2}\frac{d^2a}{dt^2},\,\,j(t)\equiv \frac{1}{aH^3}\frac{d^3a}{dt^3},\\&
s(t)\equiv \frac{1}{aH^4}\frac{d^4a}{dt^4},\,\,l(t)\equiv \frac{1}{aH^5}\frac{d^5a}{dt^5}.\\&
\end{aligned}
\end{equation}
We can expand $D_{\rm M}(z)$ up to the fifth order,

\begin{equation}
D_{\rm M}(z)=\frac{c}{H_0}\sum_{i=1}d_iz^i,
\label{Dm}
\end{equation}
where

\begin{equation}
\begin{aligned}
&d_1=1,\\&
d_2=-\frac{1}{2}(1+q_0),\\&
d_3=\frac{1}{6}(2+4q_0+3q_0^2-j_0+\Omega_{\rm K}),\\&
d_4=\frac{1}{24}[-6-18q_0+27q_0^2+15q_0^3+j_0(9+10q_0)+s_0\\&
-6\Omega_{\rm K}(1+q_0)],\\&
d_5=\frac{1}{120}[24+10j_0^2-l_0+96q_0+216q_0^2+240q_0^3+105q_0^4\\&
-j_0(72+160q_0+105q_0^2)-16s_0-15q_0s_0+5\Omega_{\rm K}(7-2j_0\\&
+14q_0+9q_0^2)+\Omega_{\rm K}^2],\\&
\end{aligned}
\end{equation}
with the subscript ``0'' denoting the values at the present time. Obviously, the $H(z)$ can be expanded directly with the cosmographic parameters, but the expansion is independent of $\Omega_{\rm K}$. In order to expand $H(z)$ with $\Omega_{\rm K}$ included and improve the usefulness of OHD, we obtain the Hubble function in terms of $\Omega_{\rm K}$ and $D_{\rm M}(z)$,

\begin{equation}
H(z,\Omega_{\rm K})=\frac{c}{\partial D_{\rm M}(z)/\partial z}\sqrt{1+\frac{H_0^2\Omega_{\rm K}}{c^2}D_{\rm M}(z)^2}.
\label{Hz}
\end{equation}

However, \citet{Cattoen&Visser2007} showed that $z$-based expansions must break down for $z > 1$. In order to avoid the convergent problem at high redshifts, improved cosmographic techniques have been proposed, namely the auxiliary $y$-variables and Pad\'e approximations. A fairly well-known $y$ variable is given by

\begin{equation}
y_1=\frac{z}{1+z},
\end{equation}
which is proposed by~\citet{Cattoen&Visser2007} and well performed all the way back to the big bang with a nice finite range $[0,1)$ for $z \in [0,\infty)$. Moreover, in the work~\citet{Aviles2012}, another parametrization

\begin{equation}
y_2=\arctan z,
\end{equation}
is adopted, which behaves smoothly with the arctangent function. It has been demonstrated that $y_2$ can give well-defined limits $[0,\frac{\pi}{2})$ in the range of $z\in [0,\infty)$~\citep{Aviles2012}.
Besides, the third parametrization

\begin{equation}
y_3=\ln(1+z),
\end{equation}
is introduced by~\citet{Semiz2015}. As a natural logarithmic function, $y_3$ tends to infinity as $z \rightarrow \infty$, but it increases slowly with the redshift growing.

In addition to the conventional methodology of cosmography applying Taylor expansions of observables, \citet{Gruber&Luongo2014} employed Pad\'e approximants which have the superior convergence properties. For a given function $f(z)$, the Pad\'e approximant of order $(m,n)$ is given by

\begin{equation}
P_{mn}(z)=\frac{a_0+a_1z+a_2z^2+\cdots+a_mz^m}{1+b_1z+b_2z^2+\cdots+b_nz^n},
\end{equation}
where $m$ and $n$ are non-negative integers, and $a_{i}$ and $b_{i}$ are constant that satisfy the conditions $P_{mn}(0)=f(0)$, $P_{mn}^{\prime}(0)=f^{\prime}(0)$,..., $P_{mn}^{m+n}=f^{m+n}(0)$. According to the plots of the transverse comoving distance $D_{\rm M}(z)$ and Hubble parameter $H(z)$ for the third-, fourth- and fifth-order Pad\'e approximants in Figure 1 of the work ~\citet{Li2020}, the $P_{12}$, $P_{22}$, $P_{13}$, $P_{32}$ expansions can give good approximations to the $\Lambda$CDM model over the redshift interval $z < 2.4$ while other expansions will diverge from $\Lambda$CDM model outside a low redshift region.

In this work, we adopt these three parameterizations of Taylor series as well as the Pad\'e approximants to reconstruct the expansion evolution of the Universe and constrain $H_0$, $\Omega_{\rm K}$ and $r_{\rm d}$. In order to achieve high accurate performance without introducing too many model parameters, the third-, fourth- and fifth-order approximants, namely $y_{i}^{(j)}$($i$=1,2,3; $j$=3,4,5) for the $y$ series and $P_{12}$, $P_{22}$, $P_{13}$, $P_{32}$ for the Pad\'e series, are used. We also utilize the Bayesian evidence method to study which cosmographic approach performs the best. More details about the explicit cosmography expressions and Bayesian evidence are reported in Appendix~\ref{A1} and~\ref{A2}.

As a guideline, the Hubble parameter in the $\Lambda$CDM+$\Omega_{\rm K}$ model is introduced,

\begin{equation}
H(z)=H_0\sqrt{\Omega_{\rm m}(1+z)^3+\Omega_{\rm K}(1+z)^2+(1-\Omega_{\rm m}-\Omega_{\rm K})},
\end{equation}
with corresponding cosmographic parameters being

\begin{equation}
q_0=\frac{3}{2}\Omega_{\rm m}+\Omega_{\rm K}-1,
\end{equation}

\begin{equation}
j_0=1-\Omega_{\rm K},
\end{equation}

\begin{equation}
s_0=1-\frac{9}{2}\Omega_{\rm m}+\Omega_{\rm K}^2-\Omega_{\rm K}(2-\frac{3}{2}\Omega_{\rm m}),
\end{equation}

\begin{equation}
l_0=1+3\Omega_{\rm m}+\frac{27}{2}\Omega_{\rm m}^2+\Omega_{\rm K}^2-\Omega_{\rm K}(2-9\Omega_{\rm m}),
\end{equation}
where $\Omega_{\rm m}$ represents the current matter density. In particular, when $\Omega_{\rm K}=0$ corresponds to the flat $\Lambda$CDM model, the terms containing $\Omega_{\rm K}$ in the above equations will disappear.

In this paper, we use the newest Pantheon+ SNe Ia sample~\citep{Brout2022,Scolnic2022}, the 14 latest BAO measurements~\citep{Alam2021}, and the 32 OHD~\citep{Wu2023} obtained from cosmic chronometer method. More details about the data and fitting methods are reported in Appendix~\ref{A3}.

\begin{table}
 \caption{\label{Priors}The priors for the model parameters.}
 \begin{center}
   \begin{tabular}{ccc}
   \hline
$\rm Parameters$ & $$ & $\rm Priors$ \\
\hline
$H_0$ & $$ & $\left[50,90\right]$ \\
$\Omega_{\rm K}$ & $$ & $\left[-0.5,0.5\right]$\\
$q_0$ & $$ & $\left[-2,0\right]$ \\
$j_0$ & $$ & $\left[-10,10\right]$ \\
$s_0$ & $$ & $\left[-70,70\right]$ \\
$l_0$ & $$ & $\left[-500,500\right]$ \\
$r_d$ & $$ & $\left[130,160\right]$ \\
\hline
\end{tabular}
\end{center}
\end{table}

\begin{table*}
 \caption{Constrained cosmographic parameters by the SNe Ia+OHD+BAO dataset under various expansion orders within the 1$\sigma$ confidence level. Here, `-' denotes there is no constraint result.}
 \label{Tabomk}
\begin{center}
\begin{tabular}{c c c c c c c c}
\hline
\hline
${\rm Model}$ & $H_0$ & $\Omega_{\rm K}$ & $q_0$ & $j_0$ & $s_0$ & $l_0$ & $r_{\rm d}$\\
   \hline
${\rm Flat\,\,\Lambda CDM}$ & $68.30\pm 1.66$ & $0$ & $-0.53\pm0.02$ & $1$ & $-0.41\pm0.06$ & $3.27\pm0.14$ & $145.70\pm 3.44$\\

${\rm \Lambda CDM+}\Omega_{\rm K}$ & $67.59\pm 1.66$ & $0.095\pm 0.064$ & $-0.48\pm0.04$ & $0.92\pm0.07$ & $-0.43\pm0.06$ & $3.07\pm0.24$ & $146.47\pm 3.46$\\

$y_1^{(3)}$ & $68.27\pm 1.66$ & $-0.123\pm 0.048$ & $-0.40\pm 0.11$ & $-0.7\pm 1.1$ & $-$ & $-$ & $144.99^{+3.14}_{-3.54}$\\

$y_1^{(4)}$ & $68.06\pm 1.66$ & $-0.047\pm 0.080$ & $-0.54^{+0.13}_{-0.18}$ & $2.3^{+3.0}_{-2.0}$ & $18.8\pm26.6$ & $-$ & $145.52\pm 3.43$\\

$y_1^{(5)}$ & $67.82\pm 1.71$ & $-0.013^{+0.084}_{-0.095}$ & $-0.42\pm0.17$ & $-0.7\pm2.8$ & $<-11.3$ & $<-98$ & $145.68\pm3.56$\\

$y_2^{(3)}$ & $69.07\pm 1.66$ & $-0.219\pm 0.056$ & $-0.45\pm 0.05$ & $0.9\pm0.1$ & $-$ & $-$ & $143.44\pm 3.34$\\

$y_2^{(4)}$ & $68.36\pm 1.69$ & $-0.040\pm 0.086$ & $-0.63\pm 0.08$ & $2.4^{+0.5}_{-0.6}$ & $2.4^{+0.4}_{-1.1}$ & $-$ & $144.96\pm3.43$\\

$y_2^{(5)}$ & $67.06\pm1.68$ & $0.029\pm0.094$ & $-0.42^{+0.14}_{-0.11}$ & $-0.2^{+1.0}_{-1.5}$ & $-2.7^{+1.6}_{-2.6}$ & $71^{+30}_{-70}$ & $147.68\pm3.61$\\

$y_3^{(3)}$ & $67.83\pm 1.71$ & $0.005\pm 0.093$ & $-0.451\pm 0.046$ & $1.0^{+0.1}_{-0.2}$ & $-$ & $-$ & $145.73\pm 3.53$\\

$y_3^{(4)}$ & $67.56\pm 1.76$ & $0.025^{+0.087}_{-0.098}$ & $-0.54\pm 0.08$ & $1.8\pm 0.6$ & $5.7^{+2.6}_{-3.9}$ & $-$ & $146.44\pm 3.70$\\

$y_3^{(5)}$ & $67.08^{+1.64}_{-1.93}$ & $-0.001\pm0.092$ & $-0.42^{+0.09}_{-0.13}$ & $0.2^{+1.5}_{-1.9}$ & $4.1\pm6.6$ & $6^{+12}_{-27}$ & $147.52\pm3.79$\\

$P_{\rm 12}$ & $67.54\pm 1.73$ & $0.032^{+0.088}_{-0.101}$ & $-0.38\pm 0.03$ & $0.4\pm0.1$ & $-$ & $-$ & $145.76^{+3.37}_{-3.75}$\\

$P_{\rm 22}$ & $67.51\pm 1.71$ & $0.043\pm0.083$ & $-0.57^{+0.10}_{-0.14}$ & $3.2^{+1.8}_{-1.2}$ & $> 27.4$ & $-$ & $146.36^{+3.36}_{-3.79}$\\

$P_{\rm 13}$ & $67.37\pm 1.75$ & $0.008^{+0.087}_{-0.099}$ & $-0.46\pm 0.05$ & $0.8\pm 0.2$ & $-0.6\pm 0.1$ & $-$ & $146.74^{+3.45}_{-3.82}$\\

$P_{\rm 32}$ & $67.64\pm1.77$ & $0.038\pm0.097$ & $-0.52^{+0.11}_{-0.13}$ & $2.2^{+1.6}_{-1.0}$ & $16.5^{+17.0}_{-8.7}$ & $>159$ & $146.01\pm3.63$\\
   \hline
  \end{tabular}
\end{center}
\end{table*}

\begin{figure*}
    \centering
    \includegraphics[scale=0.5]{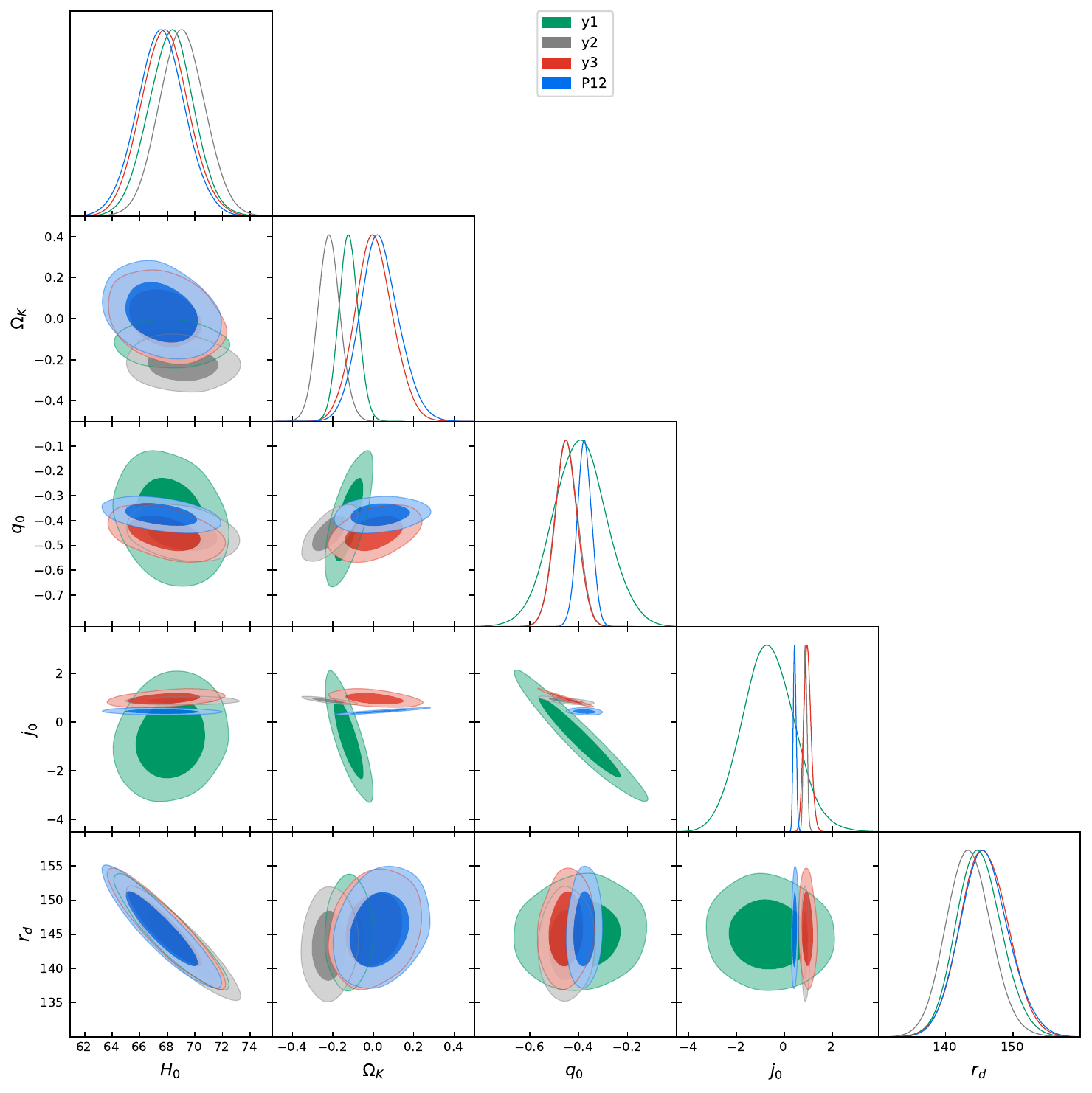}
    \caption{One-dimensional and two-dimensional marginalized distributions with 1$\sigma$ and 2$\sigma$ contours for the selected parameters under the third order.}
\label{Figomk1}
\end{figure*}

\begin{figure*}
    \centering
    \includegraphics[scale=0.5]{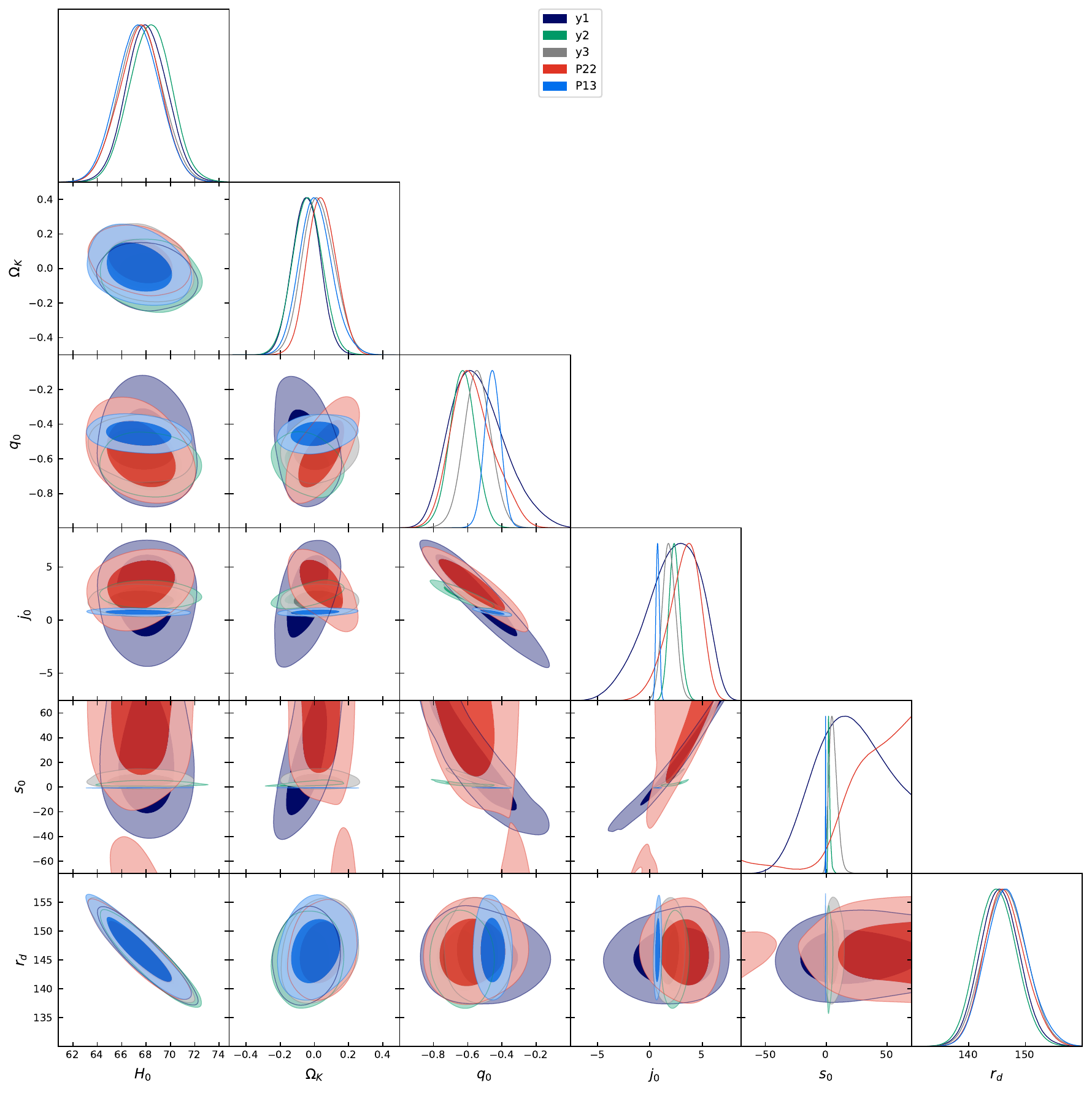}
    \caption{One-dimensional and two-dimensional marginalized distributions with 1$\sigma$ and 2$\sigma$ contours for the selected parameters under the fourth order.}
\label{Figomk2}
\end{figure*}

\begin{figure*}
    \centering
    \includegraphics[scale=0.5]{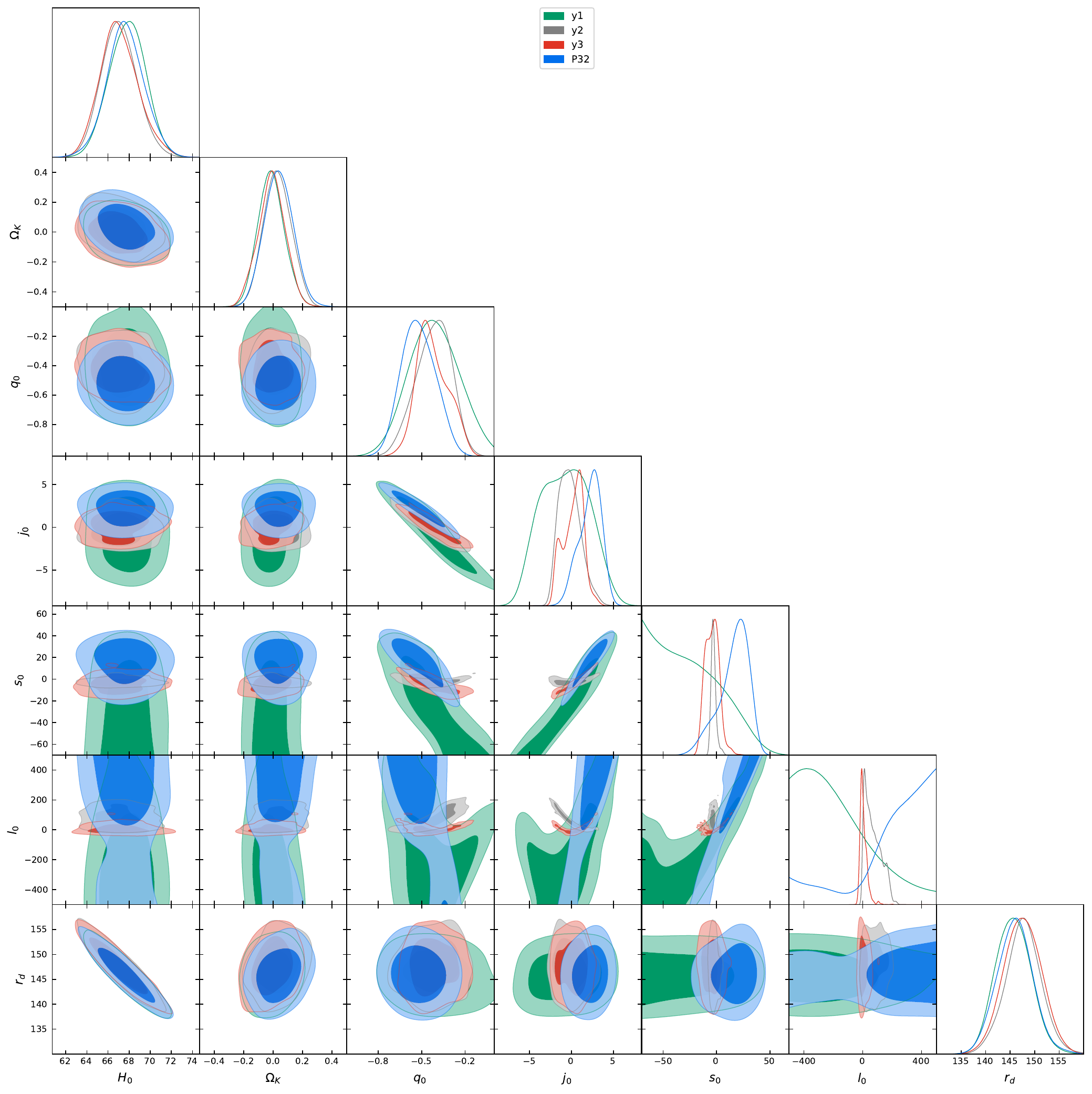}
    \caption{One-dimensional and two-dimensional marginalized distributions with 1$\sigma$ and 2$\sigma$ contours for the selected parameters under the fifth order.}
\label{Figomk3}
\end{figure*}

\begin{figure*}
    \centering
    \includegraphics[scale=0.55]{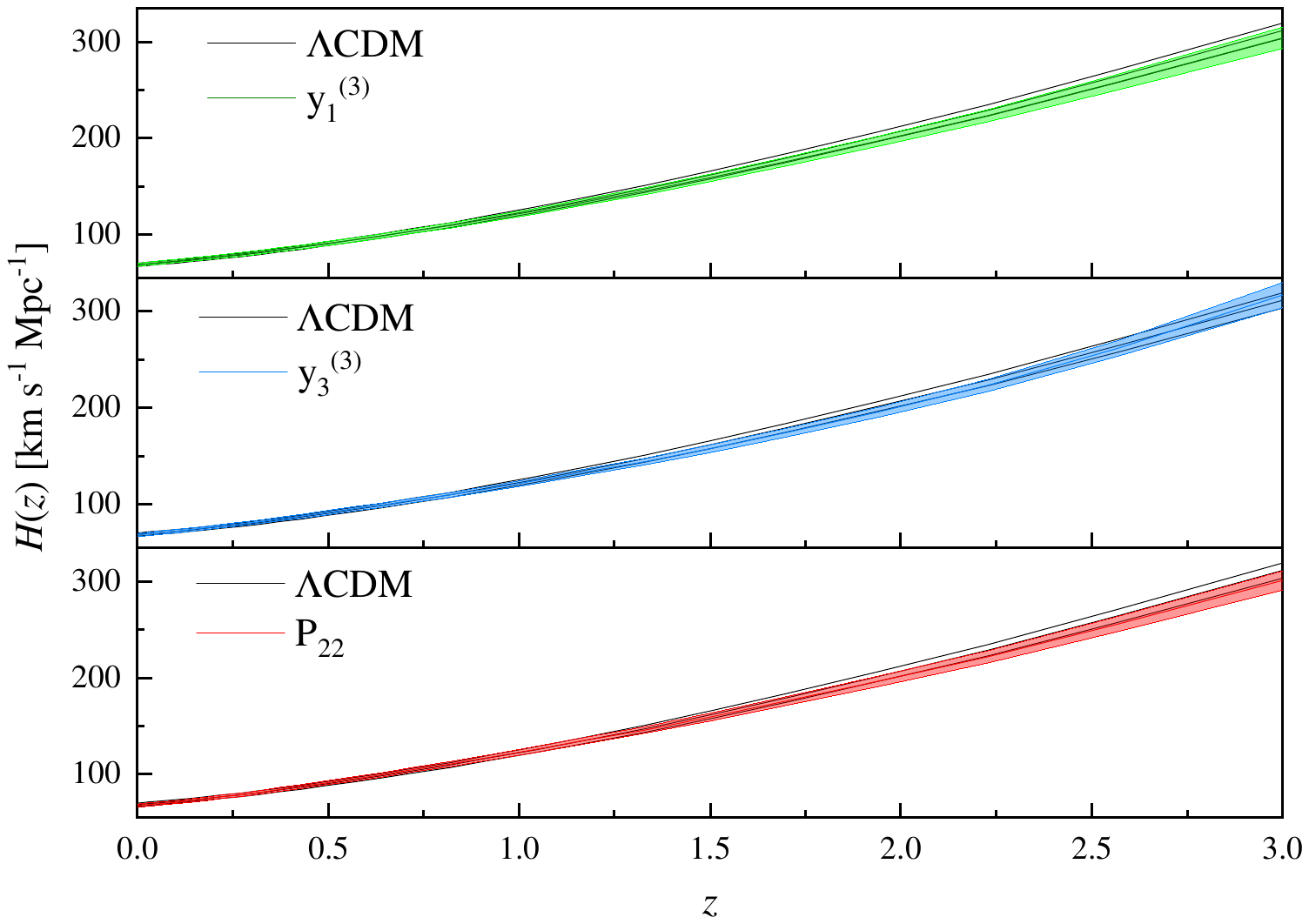}
    \caption{The reconstructed evolutions of $H(z)$ with the flat $\Lambda$CDM, $y_1^{(3)}$, $y_3^{(3)}$ and $P_{\rm 22}$ models. The black lines show the center values and 1$\sigma$ errors of $H(z)$ for flat $\Lambda$CDM model. The dark green, blue and red lines with light bands show the center values and 1$\sigma$ errors of $H(z)$ for $y_1^{(3)}$, $y_3^{(3)}$ and $P_{\rm 22}$ models, respectively.}
\label{fighz}
\end{figure*}

\begin{table}
 \caption{\label{TabBayes}The values of ln$ B_{ij}$ computed for the selected cosmography model $\mathcal{M}_i$, where the reference scenario is the flat $\Lambda$CDM model ($\mathcal{M}_j$).}
 \begin{center}
   \begin{tabular}{c|c|c}
   \hline
${\rm Model}$ & $$ln$ B_{ij}$ & ${\rm Evidence\,\,against\,\,flat \,\,\Lambda CDM}$ \\
\hline
${\rm \Lambda CDM+}\Omega_{\rm K}$ & $-0.65\pm0.16$ & ${\rm Weak}$ \\
$y_1^{(3)}$ & $-2.53\pm0.16$ & ${\rm Definite/positive}$ \\
$y_1^{(4)}$ & $-2.66\pm0.16$ & ${\rm Definite/positive}$ \\
$y_1^{(5)}$ & $-2.60\pm0.17$ & ${\rm Definite/positive}$ \\
$y_2^{(3)}$ & $-7.62\pm0.16$ & ${\rm Very\,strong}$ \\
$y_2^{(4)}$ & $-5.32\pm0.17$ & ${\rm Very\,strong}$ \\
$y_2^{(5)}$ & $-5.26\pm0.18$ & ${\rm Very\,strong}$ \\
$y_3^{(3)}$ & $-2.43\pm0.16$ & ${\rm Definite/positive}$ \\
$y_3^{(4)}$ & $-3.45\pm0.17$ & ${\rm Strong}$ \\
$y_3^{(5)}$ & $-4.40\pm0.17$ & ${\rm Strong}$ \\
$P_{\rm 12}$ & $-3.70\pm0.16$ & ${\rm Strong}$ \\
$P_{\rm 22}$ & $-1.92\pm0.16$ & ${\rm Definite/positive}$ \\
$P_{\rm 13}$ & $-3.97\pm0.17$ & ${\rm Strong}$ \\
$P_{\rm 32}$ & $-5.57\pm0.18$ & ${\rm Very\,Strong}$ \\
\hline
\end{tabular}
\end{center}
\end{table}


\section{Constraint Results}\label{S3}
We use CosmoMC~\citep{Lewis&Bridle2002} together with a nested sampling plug-in, namely PolyChord~\citep{Handley2015a,Handley2015b} which enables the computation of the Bayesian evidence, to study four cosmography models. With imposing uniform priors on the free parameters as listed in Table~\ref{Priors}, we obtain the constraint results by numerical calculations, and summarize the mean and 68.3\% confidence limits of cosmological parameters in Table~\ref{Tabomk}. And, the likelihood distributions of $y_1$, $y_2$, $y_3$ and Pad\'e approximations from the third order to the fifth order are shown in Figure~\ref{Figomk1}$-$~\ref{Figomk3}, respectively. Furthermore, the logarithmic values of the Bayes factor ln$B_{ij}$ for the $\Lambda$CDM+$\Omega_{\rm K}$ and the four cosmography models are given in Table~\ref{TabBayes} with the flat $\Lambda$CDM model as the reference model.

First, we focus on the Hubble constant $H_0$ and spatial curvature parameter $\Omega_{\rm K}$. From Table~\ref{Tabomk} we can see that the values of $H_0$ obtained from the four cosmographic approaches are consistent with the P18 value at 1$\sigma$ confidence level (CL), and lower than the R22 value by 2.3$\sim$3.0$\sigma$. Meanwhile, they are in good agreement with that obtained from the flat $\Lambda$CDM and $\Lambda$CDM$+\Omega_{\rm K}$ models. Additionally, it is worth noting that under the $\Lambda$CDM$+\Omega_{\rm K}$ model and all cosmography models, a flat Universe is preferred by the present observations except for the $y_1^{(3)}$ and $y_2^{(3)}$ expansions in which a closed Universe is preferred at more than 2$\sigma$ confidence level (CL). Meanwhile, from the $H_0-\Omega_{\rm K}$ plane in Figure~\ref{Figomk1}$-$~\ref{Figomk3}, we can find that there is a visible anti-correlation between $H_0$ and $\Omega_{\rm K}$, indicating that a larger $H_0$ can be obtained from a smaller $\Omega_{\rm K}$, and in all the $y$ approximations, the lower the expansion order, the smaller the constraint $\Omega_{\rm K}$.

Second, compared with $r_{\rm d}=147.05\pm0.3$ Mpc reported by Planck 2018, our constraints on $r_{\rm d}$ using all cosmographic approaches are in agreement with Planck 2018 within 1$\sigma$ CL. Since the values of $r_{\rm d}$ are essentially the same including the mean value and the uncertainty in four cosmography models, from Table~\ref{Tabomk} and Figure~\ref{Figomk1}$-$~\ref{Figomk3}, it can be concluded that the sound horizon $r_{\rm d}$ is not sensitive to the parameterization and truncation of the expansions. In addition, from the $H_0-r_{\rm d}$ contours, we can see that there is an obvious anti-correlation between $r_{\rm d}$ and $H_0$. Therefore, if the expansion evolution of the Universe is determined, a smaller $r_{\rm d}$ would result in a larger $H_0$, which can also be seen from Equation~(\ref{rd}).

Third, we analyze the constraints on the cosmological parameters $q_0$, $j_0$, $s_0$ and $l_0$. From Table~\ref{Tabomk}, it can be seen that the observations could provide a good constraint on the lower-order expansions, within which the values of $q_0$ and $j_0$ are basically consistent with that derived with the flat $\Lambda$CDM model, except the (1,2) order of Pad\'e approximant. However, due to the complexity of high expansion orders and the insufficiency of high-redshift data, there are large uncertainties in the constraints of $s_0$ and $l_0$ by all the cosmography techniques. This suggests that new probes into the evolution of the Universe are needed to break the degeneracies of these parameters. Meanwhile, one can see from Table~\ref{Tabomk} that the uncertainties on the parameters of $H_0$, $\Omega_{\rm K}$, and $r_{\rm d}$ remain roughly the same as one increases the expansion order, which is similar to the results obtained by~\citet{Li2020}. This is mainly because the degeneracy between the parameters of $H_0$, $\Omega_{\rm K}$, and $r_{\rm d}$, and the cosmographic parameters decreases with the expansion order increases as shown in Figures~\ref{Figomk1}$-$\ref{Figomk3}, and thus the constraint ability of observational data on the parameters of $H_0$, $\Omega_{\rm K}$, and $r_{\rm d}$ is not substantially weakened with the increase of the number of model parameters.

Fourth, we apply the Bayesian evidences with the flat $\Lambda$CDM as the reference model to determine the most favored cosmography model for the joint datasets. From Table~\ref{TabBayes}, one can see that all the ln$B_{ij}$ values in this table are negative, indicating that the flat $\Lambda$CDM is the most preferred model by the observations, and the $\Lambda$CDM+$\Omega_{\rm K}$ model is also more favored than the four cosmography models. Furthermore, according to the Jeffreys scale list in Table~\ref{Tabrule}, the $\lvert\mathrm{lnB}_{ij}\rvert$ values of five models, i.e. all expansions of $y_1$, the third order of $y_3$ and the (2,2) order of Pad\'e approximant, are smaller than 3, indicating that those models could fit the observations well. Especially, $y_1^{(3)}$, $y_3^{(3)}$ and $P_{\rm 22}$ are the top three models that are most supported by the observations. Meanwhile, we find that all the $y_2$ expansions have the largest values of $\lvert\mathrm{lnB}_{ij}\rvert$ greater than 5, which means these kinds of expansions are strongly disfavored by the observational data, and can therefore be ruled out. In addition, we derive the evolutions of $H(z)$ within the $y_1^{(3)}$, $y_3^{(3)}$ and $P_{\rm 22}$ models, and plot them combining the one obtained from the flat $\Lambda$CDM model for a comparison in Figure~\ref{fighz}. One can see that the evolutions of $H(z)$ are consistent well with the prediction of flat $\Lambda$CDM model at the lower redshift~($z<1\sim1.5$). And although the evolutions of $H(z)$ from the cosmography models have some deviations from the flat $\Lambda$CDM model at the higher redshift, it is still in agreement with the flat $\Lambda$CDM model within 1$\sigma$ CL.

Lastly, we compare our results with the previous works on the constraints of $H_0$, $\Omega_{\rm K}$ and $r_{\rm d}$ by model-independent methods. Adopting Taylor series in terms of $y_1$ and the Pad\'e approximants along with the SNe Ia+OHD+BAO data, \citet{Li2020} showed that the tension level of $H_0$ has less than 2$\sigma$ significance between different approximations and the local distance ladder determination while our constraint results show a tension of more than 2$\sigma$ using the latest data. Furthermore, their constraint results on $\Omega_{\rm K}$ preferred a closed Universe under all the approximations, whereas we find the newest datasets prefer a flat Universe with a higher precision. Therefore, the latest BAO measurements and Pantheon+ sample data can improve the accuracy of constraints on $H_0$ and $\Omega_{\rm K}$. Moreover, through Bayesian evidence, \citet{Li2020} found that the $y_1^{(3)}$ has weak evidence against $\Lambda$CDM+$\Omega_{\rm K}$ model and $P_{\rm 22}$ has positive evidence, while our constraint results show that $y_3^{(3)}$ behaves the best in all series of $y$-variables with the updated BAO data and Pantheon+ SNe Ia sample. In addition, assuming a spatially flat Universe, \citet{Zhang&Huang2021} reconstructed $H(z)$ in cubic expansion and polynomial expansion respectively, and constrained Hubble constant and sound horizon with the joint data of SNe Ia, BAO measurements, OHD, and GW data. And they found $H_0$ and $r_{\rm d}$ were consistent with the estimate derived from Planck 2018 data, which is similar to our constraint results. Likewise, \citet{Cao2022} combined 55 simulated SGLTD data and 1000 simulated GW data, and obtained high-precision constraint results $H_0=73.65\pm0.33$ km $\mathrm{s}^{-1} \mathrm{Mpc}^{-1}$ and $\Omega_{\rm K}=0.008\pm0.038$. Although those results showed a strong constraint on $H_0$ comparing to our constraint results, those works use simulated data, whereas we use actual observational data.



\section{Conclusion and Discussions}\label{S4}

The Hubble tension problem can be relaxed to some extent by introducing $\Omega_{\rm K}$ or any cosmological model that assumes a small $r_{\rm d}$ in the early Universe. In this paper, using the Pantheon+ sample of SNe Ia which significantly enlarges the origin Pantheon sample size from 1048 to 1701, the 14 latest BAO measurements, and 32 OHD data, we constrain Hubble constant $H_0$ while leaving spatial curvature parameter $\Omega_{\rm K}$ and sound horizon $r_{\rm d}$ completely free. In order to be model-independent, four cosmography approaches are taken, namely the Taylor expansions in terms of $y_1=z/(1+z)$, $y_2=\arctan z$, $y_3=\ln (1+z)$ and the Pad\'e approximants. We expand the transverse comoving distance and Hubble parameter with these four variables, and obtain the constraint results on $H_0$, $\Omega_{\rm K}$, and $r_{\rm d}$ from the joint datasets. Then we adopt Bayesian evidence to determine which cosmography model gives the best approximation.

From the constraint results, we find that the newest Pantheon+ sample and updated BAO measurements could tighten the constraints on $H_0$ and $\Omega_{\rm K}$ obviously. And the values of $H_0$ obtained from the four cosmography models are in good agreement with P18 and lower than R22 by 2.3$\sim$3.0$\sigma$. Meanwhile, a flat Universe is preferred by the observations under the $\Lambda$CDM+$\Omega_{\rm K}$ model and all cosmography approaches, except for $y_1^{(3)}$ and $y_2^{(3)}$. In addition, the constraints on $r_{\rm d}$ are consistent with the estimate derived from the Planck CMB data within 1$\sigma$ under all expansions, suggesting that $r_{\rm d}$ is not sensitive to the approximations parameterizations. Furthermore, we find that there are anti-correlations between $H_0$ and $\Omega_{\rm K}$ or $H_0$ and $r_{\rm d}$, indicating smaller $\Omega_{\rm K}$ or $r_{\rm d}$ would be helpful in alleviating the $H_0$ tension. Moreover, according to the Bayesian evidence, the $\Lambda$CDM is still the most favored model by the joint datasets. As for the cosmography models, the Pad\'e approximant of order (2,2), the third order of $y_3$ and $y_1$ are the top three models that fit the datasets best, but the Taylor series in terms of $y_2$ are essentially excluded by cosmological observations.

As the final remarks, the release of $H_0$ tension in this paper is mainly caused by the increase in its uncertainty. If we take into account the full covariance of OHD~\citep{Moresco2020,Moresco2022}, the $H_0$ tension with R22 will decrease from 2.3$\sim$3.0$\sigma$ to 1.0$\sim$1.9$\sigma$ due to the further increased uncertainty. At the same time, the cosmographic parameters can not be constrained effectively when expanding up to high order. This suggests that more high-precision data and additional probes are needed to measure the cosmic expansion due to the degeneracy with $\Omega_{\rm K}$ and cosmographic parameters. In the future, with the space-based GW detectors such as DECIGO~\citep{Seto2001,Kawamura2011}, Taiji~\citep{Hu&Wu2017}, TianQin~\citep{Luo2016} and ET (Einstein Telescope)~\citep{Punturo2010} exploring more GW from BNS systems with EM counterparts, the constraints on $H_0$ and $\Omega_{\rm K}$ will be more accurate and tight.


\acknowledgments
We thank the anonymous reviewer for the insightful suggestions and comments. We also appreciate Professor Puxun Wu for the useful discussions. This work was supported by the National Natural Science Foundation of China under grants Nos. 12174227, 11974219, 11505004, and 11865018, the University Scientific Research Project of Anhui Province under grant No. 2022AH051634, and the Natural Science Foundation of Anhui Province under grant No. 1508085QA17.

\bibliography{}

\appendix
\section{cosmographic expansions}\label{A1}

\;\;In this section, all the cosmographic expansions up to the fifth order which we adopted in this study are reported.

(i)  the transverse comoving distance within the $y_1$-variable model:

\begin{equation}
\begin{aligned}
D_{\rm M}(y_1)=&\frac{c}{H_0}[y_{\rm 1}+\frac{1}{2}(1-q_0)y_{\rm 1}^2+\frac{1}{6}(2-j_0-2q_0+3q_0^2+\Omega_{\rm K})y_{\rm 1}^3\\&+\frac{1}{24}(6-3j_0-6q_0+10j_0q_0+9q_0^2-15q_0^3+s_0+6\Omega_{\rm K}-6q_0\Omega_{\rm K})y_{\rm 1}^4\\&+\frac{1}{120}(24-12j_0+10j_0^2-l_0-24q_0+40j_0q_0+36q_0^2-105j_0q_0^2-60q_0^3\\&+105q_0^4+4s_0-15q_0s_0+35\Omega_{\rm K}-10j_0\Omega_{\rm K}-50q_0\Omega_{\rm K}+45q_0^2\Omega_{\rm K}+\Omega_{\rm K}^2)y_{\rm 1}^5]
\end{aligned}
\end{equation}

(ii) the transverse comoving distance within the $y_2$-variable model:

\begin{equation}
\begin{aligned}
D_{\rm M}(y_2)=&\frac{c}{H_0}[y_{\rm 2}+\frac{1}{2}(-1-q_0)y_{\rm 2}^2+\frac{1}{6}(4-j_0+4q_0+3q_0^2+\Omega_{\rm K})y_{\rm 2}^3\\&+\frac{1}{24}(-14+9j_0-26q_0+10j_0q_0-27q_0^2-15q_0^3+s_0-6\Omega_{\rm K}-6q_0\Omega_{\rm K})y_{\rm 2}^4\\&+\frac{1}{120}(80-92j_0+10j_0^2-l_0+176q_0-160j_0q_0+276q_0^2-105j_0q_0^2+240q_0^3\\&+105q_0^4-16s_0-15q_0s_0+55\Omega_{\rm K}-10j_0\Omega_{\rm K}+70q_0\Omega_{\rm K}+45q_0^2\Omega_{\rm K}+\Omega_{\rm K}^2)y_{\rm 2}^5]
\end{aligned}
\end{equation}

(iii) the transverse comoving distance within the $y_3$-variable model:

\begin{equation}
\begin{aligned}
D_{\rm M}(y_3)=&\frac{c}{H_0}[y_{\rm 3}-\frac{1}{2}q_0y_{\rm 3}^2+\frac{1}{6}(-j_0+q_0+3q_0^2+\Omega_{\rm K})y_{\rm 3}^3\\&+\frac{1}{24}(3j_0-q_0+10j_0q_0-9q_0^2-15q_0^3+s_0-6q_0\Omega_{\rm K})y_{\rm 3}^4\\&+\frac{1}{120}(-7j_0+10j_0^2-l_0+q_0-60j_0q_0+21q_0^2-105j_0q_0^2+90q_0^3\\&+105q_0^4-6s_0-15q_0s_0 -10j_0\Omega_{\rm K}+10q_0\Omega_{\rm K}+45q_0^2\Omega_{\rm K}+\Omega_{\rm K}^2)y_{\rm 3}^5]
\end{aligned}
\end{equation}

(iv) $P_{\rm 12}$ approximation of the transverse comoving distance:

\begin{equation}
\begin{aligned}
&P_{\rm12}=\frac{c}{H_0}\frac{z}{1+b_1z+b_2z^2}\\&
b_1=\frac{1}{2}(1+q_0)\\&
b_2=\frac{1}{12}(-1+2j_0-2q_0-3q_0^2-2\Omega_{\rm K})\\&
\end{aligned}
\end{equation}

(v) $P_{\rm 13}$ approximation of the transverse comoving distance:

\begin{equation}
\begin{aligned}
&P_{\rm13}=\frac{c}{H_0}\frac{z}{1+b_1z+b_2z^2+b_3z^3}\\&
b_1=\frac{1}{2}(1+q_0)\\&
b_2=\frac{1}{12}(-1+2j_0-2q_0-3q_0^2-2\Omega_{\rm K})\\&
b_3=\frac{1}{24}(1-5j_0+3q_0-6j_0q_0+8q_0^2+6q_0^3-s_0+2\Omega_{\rm K}+2q_0\Omega_{\rm K})\\&
\end{aligned}
\end{equation}

(vi) $P_{\rm 22}$ approximation of the transverse comoving distance:

\begin{equation}
\begin{aligned}
&P_{\rm22}=\frac{c}{H_0}\frac{z+\frac{1}{2}\frac{a_2}{c_1}z^2}{1+\frac{1}{2}\frac{b_1}{c_1}z+\frac{1}{12}\frac{b_2}{c_1}z^2}\\&
a_2=-1+5j_0-3q_0+6j_0q_0-8q_0^2-6q_0^3+s_0-2\Omega_{\rm K}-2q_0\Omega_{\rm K}\\&
b_1=-2+7j_0-6q_0+8j_0q_0-13q_0^2-9q_0^3+s_0-4\Omega_{\rm K}-4q_0\Omega_{\rm K}\\&
b_2=2-11j_0-4j_0^2+8q_0-25j_0q_0+23q_0^2-6j_0q_0^2+30q_0^3-3s_0\\&-3q_0s_0+2\Omega_{\rm K}+8j_0\Omega_{\rm K}+4q_0\Omega_{\rm K}-6q_0^2\Omega_{\rm K}-4\Omega_{\rm K}^2\\&
c_1=-1+2j_0-2q_0-3q_0^2-2\Omega_{\rm K}\\&
\end{aligned}
\end{equation}

(vii) $P_{\rm 32}$ approximation of the transverse comoving distance:

\begin{equation}
\begin{aligned}
&P_{\rm32}=\frac{c}{H_0}\frac{z+\frac{1}{10}\frac{a_2}{c_1}z^2+\frac{1}{60}\frac{a_3}{c_1}z^3}{1+\frac{1}{5}\frac{b_1}{c_1}z+\frac{1}{20}\frac{b_2}{c_1}z^2}\\&
a_2=14-137j_0-10j_0^2-6l_0+70q_0-472j_0q_0-20j_0^2q_0-6l_0q_0+277q_0^2-495j_0q_0^2\\&
+551q_0^3-150j_0q_0^3+465q_0^4+135q_0^5-61s_0-10j_0s_0-116q_0s_0-45q_0^2s_0+20\Omega_{\rm K}\\&
+50j_0\Omega_{\rm K}+60q_0\Omega_{\rm K}+60j_0q_0\Omega_{\rm K}+10q_0^2\Omega_{\rm K}-30q_0^3\Omega_{\rm K}+10s_0\Omega_{\rm K}-34\Omega_{\rm K}^2-34q_0\Omega_{\rm K}^2\\&
a_3=4-60j_0+9j_0^2-80j_0^3-6l_0+12j_0l_0+24q_0-276j_0q_0-12l_0q_0+120q_0^2-468j_0q_0^2\\&
+180j_0^2q_0^2-18l_0q_0^2+320q_0^3-330j_0q_0^3+423q_0^4-270j_0q_0^4+270q_0^5+135q_0^6-36s_0\\&
-18j_0s_0-102q_0s_0-60j_0q_0s_0-78q_0^2s_0-15s_0^2+18\Omega_{\rm K}-144j_0\Omega_{\rm K}+120j_0^2\Omega_{\rm K}\\&
-12l_0\Omega_{\rm K}+72q_0\Omega_{\rm K}-360j_0q_0\Omega_{\rm K}+252q_0^2\Omega_{\rm K}-420j_0q_0^2\Omega_{\rm K}+360q_0^3\Omega_{\rm K}+270q_0^4\Omega_{\rm K}\\&
-72s_0\Omega_{\rm K}-60q_0s_0\Omega_{\rm K}+6\Omega_{\rm K}^2-12j_0\Omega_{\rm K}^2+12q_0\Omega_{\rm K}^2+18q_0^2\Omega_{\rm K}^2-28\Omega_{\rm K}^3\\&
b_1=12-96j_0-15j_0^2-3l_0+60q_0-326j_0q_0-20j_0^2q_0-3l_0q_0+216q_0^2-325j_0q_0^2\\&
+408q_0^3-90j_0q_0^3+330q_0^4+90q_0^5-38s_0-5j_0s_0-73q_0s_0-30q_0^2s_0+15\Omega_{\rm K}\\&
+45j_0\Omega_{\rm K}+45q_0\Omega_{\rm K}+50j_0q_0\Omega_{\rm K}-30q_0^3\Omega_{\rm K}+5s_0\Omega_{\rm K}-27\Omega_{\rm K}^2-27q_0\Omega_{\rm K}^2\\&
b_2=12-132j_0-37j_0^2-40j_0^3-8l_0+4j_0l_0+72q-0-596j_0q_0-100j_0^2q_0\\&
-16l_0q_0+312q_0^2-898j_0q_0^2+40j_0^2q_0^2-12l_0q_0^2+768q_0^3-510j_0q_0^3+927q_0^4\\&
-180j_0q_0^4+510q_0^5+135q_0^6-68s_0-26j_0s_0-196q_0s_0-40j_0q_0s_0-162q_0^2s_0\\&
-30q_0^3s_0-5s_0^2+16\Omega_{\rm K}^2+32j_0\Omega_{\rm K}+80j_0^2\Omega_{\rm K}-4l_0\Omega_{\rm K}+64q_0\Omega_{\rm K}+60j_0q_0\Omega_{\rm K}\\&
+64q_0^2\Omega_{\rm K}-120j_0q_0^2\Omega_{\rm K}+60q_0^4\Omega_{\rm K}-4s_0\Omega_{\rm K}-32\Omega_{\rm K}^2-44j_0\Omega_{\rm K}^2-64q_0\Omega_{\rm K}^2+12q_0^2\Omega_{\rm K}^2+4\Omega_{\rm K}^3\\&
c_1=2-11j_0-4j_0^2+8q_0-25j_0q_0+23q_0^2-6j_0q_0^2+30q_0^3+9q_0^4-3s_0-3q_0s_0\\&
+2\Omega_{\rm K}+8j_0\Omega_{\rm K}+4q_0\Omega_{\rm K}-6q_0^2\Omega_{\rm K}-4\Omega_{\rm K}^2
\end{aligned}
\end{equation}

\section{Bayesian evidence method}\label{A2}

Bayesian evidence provides a statistical way to evaluate the performances of the cosmography models. For a given model $\mathcal{M}$ with parameter space $\theta$ and specific observational data $d$, the Bayesian evidence $E$ is defined as

\begin{equation}
E=p(d|\mathcal{M})=\int p(d|\theta,\mathcal{M})\pi(\theta|\mathcal{M})d\theta,
\end{equation}
where $\pi(\theta|\mathcal{M})$ is the prior of $\theta$ in model $\mathcal{M}$, and $p(d|\theta,\mathcal{M})$ is the likelihood. Then, for the two models $\mathcal{M}_i$ and $\mathcal{M}_j$, combing the Bayes theorem,  the posterior probability is

\begin{equation}
\frac{p(\mathcal{M}_i|d)}{p(\mathcal{M}_j|d)}=\frac{p(d|\mathcal{M}_i)}{p(d|\mathcal{M}_j)}\frac{\pi(\mathcal{M}_i)}{\pi(\mathcal{M}_j)}.
\end{equation}

The ratio between posterior probabilities leads to the definition of the Bayes factor $B_{ij}$, which is written in a logarithmic scale as
\begin{equation}
\mathrm{ln}\,B_{ij}=\mathrm{ln}\,\frac{p(d|\mathcal{M}_i)}{p(d|\mathcal{M}_j)}=\mathrm{ln}\,p(d|\mathcal{M}_i)-\mathrm{ln}\,p(d|\mathcal{M}_j).
\end{equation}
Then, one can determine the strength of the preference for one of the competing models over the other by means of the Jeffreys scale listed in Table~\ref{Tabrule}~\citep{Trotta2008,Kass1995}.

\begin{table}
 \caption{\label{Tabrule}Revised Jeffreys scale quantifying the observational viability of any model $\mathcal{M}_i$ compared with some reference model $\mathcal{M}_j$.}
 \begin{center}
   \begin{tabular}{c|c}
   \hline
$$ln$B_{ij}$ & ${\rm Strength\,\,of\,\,evidence\,\,for\,\,model\,\,\mathcal{M}_i}$\\
\hline
$0<|$ln$ B_{ij}|<1$ & ${\rm Weak}$\\
$1<|$ln$ B_{ij}|<3$ & ${\rm Definite/positive}$\\
$3<|$ln$ B_{ij}|<5$ & ${\rm Strong}$\\
$|$ln$ B_{ij}|>5$ & ${\rm Very\,strong}$\\
\hline
\end{tabular}
\end{center}
\end{table}

\section{data and fitting methods}\label{A3}

In this section, we introduce the observational data and fitting methods used in our work.

\subsection{\rm SNe Ia}
As the standard candle, SNe Ia is widely used to measure the cosmological luminosity distance. In our analysis, the largest SNe Ia sample Pantheon+ is used. Pantheon+ consists of 1701 light curves of 1550 spectroscopically confirmed SNe Ia across 18 different surveys, and covers the redshift range $z\in[0.00122,2.26137]$. The observed distance modulus of each SNe Ia in this sample is defined as

\begin{equation}
\mu_{\rm obs}=m^\ast_{\rm B} + \alpha X_1 - \beta\mathcal{C} - M_{\rm_B} - \delta_{\rm_{bias}} + \delta_{\rm_{host}}\,,
\end{equation}
Here, $m^\ast_{\rm B}$ is the observed peak magnitude in rest frame B-band, $X_1$ is the time stretching of the light-curve, $\mathcal{C}$ is the SNe Ia color at maximum brightness, $M_{\rm_B}$ is the absolute magnitude, $\alpha$, $\beta$ are two nuisance parameters, which should be fitted simultaneously with the cosmological parameters. And $\delta_{\rm_{bias}}$ is a correction term to account for selection biases, $\delta_{\rm_{host}}$ is the luminosity correction for residual correlations between the standardized brightness of an SNe Ia and the host-galaxy mass. In order to avoid the dependence of the nuisance parameters on the cosmological model, \citet{Kessler&Scolnic2017} proposed a new method called BEAMS with bias corrections (BBC) to calibrate the SNe Ia, and the corrected apparent magnitude
$m^\ast_{\rm B,corr}=m^\ast_{\rm B} + \alpha X_1 - \beta\mathcal{C} - M_{\rm_B} - \delta_{\rm_{bias}} + \delta_{\rm_{host}}$ for all the SNe Ia is reported in~\citet{Popovic2021}. Then the observed distance modulus is rewritten as

\begin{equation}
\mu_{\rm obs}=m^\ast_{\rm B,corr} - M_{\rm_B}\,,
\end{equation}

On the other hand, introducing the Hubble-free luminosity $d_{\rm_L}(z)=H_0D_{\rm_L}(z)$, the theoretical value of distance modulus can be derived from

\begin{equation}
\mu_{\rm th}(z)=5\mathrm{log}(d_{\rm L})+\mu_0\,,
\end{equation}
where $\mu_{\rm 0}=42.38-5\mathrm{log}h$ with $h=H_0/100$ km $\mathrm{s}^{-1} \mathrm{Mpc}^{-1}$. Therefore, the $\chi^2$ function for the Pantheon+ SNe Ia data can be written as

\begin{equation}
\chi^2_{\rm SNIa}\equiv {\boldsymbol{\Delta \mu}}^{\rm T}\cdot \mathbf{Cov}^{-1}\cdot \boldsymbol{\Delta\mu}\,,
\end{equation}
where $\Delta\mu_{\rm i}=m^\ast_{\rm B,corr}-5\mathrm{log}d_{\rm L}(z)-\mathcal{N}$, and $\mathbf{Cov}$ is the covariance matrix, respectively. Here, $\mathcal{N}=M_{\rm B}+\mu_0$, which can be marginalized over analytically with the method proposed in~\citet{Conley2011}.
Finally, the $\chi^2$ function is rewritten as

\begin{equation}
\chi^2_{\rm SNIa,marg}=a-\frac{b^2}{f}+\ln\frac{f}{2\pi}
\end{equation}
where $a\equiv\boldsymbol{\Delta m}^{\rm T}\cdot \mathbf{Cov}^{-1}\cdot \boldsymbol{\Delta m}$, $b\equiv{\boldsymbol{\Delta m}}^{\rm T}\cdot \mathbf{Cov}^{-1}\cdot \bf{1}$, $f\equiv{\bf 1}^{\rm T}\cdot {\mathbf{Cov}}^{-1}\cdot {\bf 1}$. Here, it should be pointed out that since that the sensitivity of peculiar velocities of SNe Ia is very large at $z < 0.01$ and thus the Hubble residual bias can not be negligible, the SNe Ia data with $z > 0.01$ is only used instead of the full sample, in our analysis. We refer the reader to the Ref.~(\citet{Brout2022}) for further details on this issue.

\subsection{\rm BAO}

BAO data have provided another way to probe the expansion rate and the large-scale properties of the Universe, which give information about the imprint in the primordial plasma. The BAO data used in this paper include the measurements from the Sloan Digital Sky Survey Data Release 7 (SDSS DR7) main galaxy sample (MGS)~\citep{Ross2015}, the SDSS-\uppercase\expandafter{\romannumeral3} BOSS DR12 galaxy sample~\citep{Alam2017}, the SDSS-\uppercase\expandafter{\romannumeral4} eBOSS DR16 LRG sample~\citep{Marin2020}, the SDSS-\uppercase\expandafter{\romannumeral4} eBOSS DR16 ELG sample~\citep{de Mattia2021}, the SDSS-\uppercase\expandafter{\romannumeral6} eBOSS DR16 quasars (QSO) sample~\citep{Neveux2020}, the eBOSS DR16 auto-correlations and cross-correlations of the Ly$\alpha$ absorption and quasars~\citep{du Mas des Bourboux2020}. We summarize these data in Table~\ref{TabBAO}.

We note that the likelihoods of MGS data, DR16 Ly$\alpha$-Ly$\alpha$ and DR16 Ly$\alpha$-QSO data used in this work cannot be well approximated by a Gaussian. Thus, their full likelihoods are used. The 4$\times$4 covariance matrix for DR12 LRG data from~\citet{Marin2020} is

\begin{equation}
\left(
 \begin{array}{cccc}
0.0286052&-0.04939281&0.01489688&-0.01387079\\
-0.04939281&0.5307187&-0.02423513&0.1767087\\
0.01489688&-0.02423513&0.04147534&-0.04873962\\
-0.01387079&0.1767087&-0.04873962&0.3268589\\
 \end{array}
\right)
\end{equation}
The 2$\times$2 covariance matrix for DR16 LRG data from~\citet{Bautista2021} is

\begin{equation}
\left(
 \begin{array}{cc}
0.1076634008565565&-0.05831820341302727\\
-0.05831820341302727&0.2838176386340292\\
 \end{array}
\right)
\end{equation}
The 2$\times$2 covariance matrix for DR16 QSO data from~\citet{Hou2021} is

\begin{equation}
\left(
 \begin{array}{cc}
0.63731604&0.1706891\\
0.1706891&0.30468415\\
 \end{array}
\right)
\end{equation}
In Table~\ref{TabBAO}, the observable $D_{\rm V}\equiv[czD_{\rm M}(z)/H(z)]^{1/3}$ is the volume average distance and $D_{\rm H}\equiv c/H(z)$ is the  Hubble distance. The parameter $r_{\rm d}$ is comoving sound horizon at the end of radiation drag epoch $z_{\rm d}$, shortly after recombination, when baryons decouple from the photons,

\begin{equation}
r_{\rm d}=\int_{z_{\rm d}}^{\infty}\frac{c_s(z)}{H(z)}dz,
\label{rd}
\end{equation}
where $c_{\rm s}(z)$ is the sound speed of the photon-baryon fluid. In this paper, the parameter $r_{\rm d}$ is treated as a free parameter.

Finally, the BAO data are combined into a $\chi^2$-statistic

\begin{equation}
\begin{aligned}
\chi^2_{\rm BAO}=&\chi^2_{\rm MGS}+\chi^2_{\rm DR12LRG}+\chi^2_{\rm DR16LRG}+\chi^2_{\rm DR16ELG}\\&
+\chi^2_{\rm DR16QSO}+\chi^2_{\rm DR16Ly\alpha-Ly\alpha}+\chi^2_{\rm DR16Ly\alpha-QSO},\\&
\end{aligned}
\end{equation}
Here, the DR12, DR16 LRG and DR16 QSO data $\chi^2_{\rm i}$ are given in the form of $\chi^2_{\rm i}=(\boldsymbol{w_{\rm i}}-\boldsymbol{d_{\rm i}})^T\cdot\mathbf{Cov_{\rm i}}^{-1}\cdot(\boldsymbol{w_{\rm i}}-\boldsymbol{d_{\rm i}})$. And, the vector $\boldsymbol{d_{\rm i}}$ is the observational data of the $i$th-type data set from Table~\ref{TabBAO}, $\boldsymbol{w_{\rm i}}$ is the prediction for these vectors in a given cosmological model, and $\mathbf{Cov_{\rm i}}$ is the covariance matrix of different BAO data set.

\begin{table*}
 \begin{center}
  \caption{BAO measurements used in our work.}
  \begin
  {tabular}{c|c|c|c|c}
  \hline
 ${\rm Data set}$ & $z_{\rm eff}$ & ${\rm Observable}$ & ${\rm Measurement}$ & ${\rm Reference}$\\
  \hline
 $\rm MGS$ & $0.15$ & $D_{\rm V}/r_{\rm d}$ & $4.466\pm 0.168$ & \citet{Ross2015}\\
 $\rm DR12LRG$ & $0.38$ & $D_{\rm M}/r_{\rm d}$ & $10.234$ & \citet{Alam2017}\\
 $$ & $0.38$ & $D_{\rm H}/r_{\rm d}$ & $24.980$ & $ $\\
 $$ & $0.51$ & $D_{\rm M}/r_{\rm d}$ & $13.366$ & $ $\\
 $$ & $0.51$ & $D_{\rm H}/r_{\rm d}$ & $22.317$ & $ $\\
 $\rm DR16LRG$ & $0.698$ &$D_{\rm M}/r_{\rm d}$ & $17.858$ & \citet{Marin2020}\\
 $$ & $0.698$ & $D_{\rm H}/r_{\rm d}$ & $19.326$ & $ $\\
 $\rm DR16ELG$ & $0.845$ & $D_{\rm V}/r_{\rm d}$ & $18.33\pm 0.62$ & \citet{de Mattia2021}\\
 $\rm DR16QSO$ & $1.48$ & $D_{\rm M}/r_{\rm d}$ & $30.6876$ & \citet{Neveux2020}\\
 $$ & $1.48$ & $D_{\rm H}/r_{\rm d}$ & $13.2609$ & $ $\\
 $\rm DR16Ly\alpha-Ly\alpha$ & $2.33$ & $D_{\rm M}/r_{\rm d}$ & $37.6\pm 1.9$ & \citet{du Mas des Bourboux2020}\\
 $$ & $2.33$ & $D_{\rm H}/r_{\rm d}$ & $8.93\pm 0.28$ & $ $\\
 $\rm DR16Ly\alpha-QSO$ & $2.33$ & $D_{\rm M}/r_{\rm d}$ & $37.3\pm 1.7$ & $ $\\
 $$ & $2.33$ & $D_{\rm H}/r_{\rm d}$ & $9.08\pm 0.34$ & $ $\\
  \hline
  \end{tabular}
  \label{TabBAO}
\end{center}
\end{table*}

\subsection{\rm OHD}

In this paper, we use the most recent 32 OHD points summarized in Table~\ref{TabOHD}. And the best-fitting parameters are obtained by minimizing this quantity

\begin{equation}
\chi^2_{\rm H(z)}=\sum_{i=1}^{32}\frac{\left(H_{\mathrm{obs},{\rm i}}-H_{\mathrm{th},{\rm i}}\right)^2}{\sigma_{H_{\rm i}}^2}\,,
\end{equation}
where $\sigma_{H_{\rm i}}^2$ is the error of the $\rm i$-th measurement.

\begin{table}[!hbp]
 \caption{The compilation of OHD (in units of km $\mathrm{s}^{-1} \mathrm{Mpc}^{-1}$) and their errors $\sigma_H$ at redshift $z$.}
  \begin{tabular}{c|c|c|c||c|c|c|c}
  \hline
 $z$ & $H(z)$ & $\sigma_H$ & ${\rm Reference}$ & $z$ & $H(z)$ & $\sigma_H$ & ${\rm Reference}$\\
  \hline
 $0.07$ & $69.0$ & $19.6$ & \citet{Zhang2014} & $0.4783$ & $80.9$ & $9.0$ & \citet{Moresco2016}\\
 $0.1$ & $69.0$ & $12.0$ & \citet{Simon2005} & $0.48$ & $97.0$ & $60.0$ & \citet{Stern2010}\\
 $0.12$ & $68.6$ & $26.2$ & \citet{Zhang2014} & $0.5929$ & $104.0$ & $13.0$ & \citet{Moresco2012}\\
 $0.17$ & $83.0$ & $8.0$ & \citet{Simon2005} & $0.6797$ & $92.0$ & $8.0$ & \citet{Moresco2012}\\
 $0.1791$ & $75.0$ & $4.0$ & \citet{Moresco2012} & $0.75$ & $98.8$ & $33.6$ & \citet{Borghi2022}\\
 $0.1993$ & $75.0$ & $5.0$ & \citet{Moresco2012} & $0.7812$ & $105.0$ & $12.0$ & \citet{Moresco2012}\\
 $0.20$ & $72.9$ & $29.6$ & \citet{Zhang2014} & $0.8754$ & $125.0$ & $17.0$ & \citet{Moresco2012}\\
 $0.27$ & $77.0$ & $14.0$ & \citet{Simon2005} & $0.88$ & $90.0$ & $40.0$ & \citet{Stern2010}\\
 $0.28$ & $88.8$ & $36.6$ & \citet{Zhang2014} & $0.9$ & $117.0$ & $23.0$ & \citet{Simon2005}\\
 $0.3519$ & $83.0$ & $14.0$ & \citet{Moresco2012} & $1.037$ & $154.0$ & $20.0$ & \citet{Moresco2012}\\
 $0.3802$ & $83.0$ & $13.5$ & \citet{Moresco2016} & $1.3$ & $168.0$ & $17.0$ & \citet{Simon2005}\\
 $0.4$ & $95.0$ & $17.0$ & \citet{Simon2005} & $1.363$ & $160.0$ & $33.6$ & \citet{Moresco2015}\\
 $0.4004$ & $77.0$ & $10.2$ & \citet{Moresco2016} & $1.43$ & $177.0$ & $18.0$ & \citet{Simon2005}\\
 $0.4247$ & $87.1$ & $11.2$ & \citet{Moresco2016} & $1.53$ & $140.0$ & $14.0$ & \citet{Simon2005}\\
 $0.4497$ & $92.8$ & $12.9$ & \citet{Moresco2016} & $1.75$ & $202.0$ & $40.0$ & \citet{Simon2005}\\
 $0.47$ & $89$ & $34$ & \citet{Ratsimbazafy2017} & $1.965$ & $186.5$ & $50.4$ & \citet{Moresco2015}\\
  \hline
  \end{tabular}
  \label{TabOHD}
\end{table}

\end{document}